\documentclass{PoS}

\usepackage{amssymb}
\usepackage{slashed}

\newcommand{\be}{\begin{equation}}
\newcommand{\en}{\end{equation}}
\newcommand{\bea}{\begin{eqnarray}}
\newcommand{\ena}{\end{eqnarray}}
\newcommand{\Det}{\hbox{Det}}
\newcommand{\hbo}{\hbox to 1 true cm {\hfill } }
\newcommand{\tr}{\hbox{tr}}

\title{Dense but confined matter as a new state in QCD}

\ShortTitle{Dense confined matter}

\author{\speaker{Kurt Langfeld}\thanks{It is a pleasure to thank Andreas Wipf
    for the collaboration on this project. }\\
        School of Computing \& Mathematics, University of Plymouth, Plymouth
        PL4 8AA, UK \\
        E-mail: \email{kurt.langfeld@plymouth.ac.uk}}

\abstract{Centre sector transitions in QCD-like theories with dynamical quark
  matter are investigated. In the hadronic phase, these transitions still take
place in the infinite volume at zero temperature limit despite of the explicit
breaking of the centre symmetry by the matter fields. This finding is
supported by simulations of the SU(2) Yang-Mills theory with Higgs matter.
The phenomenological impact of the centre sector
transitions for dense but confined matter is explained: centre dressed quarks
acquire Bose statistics and form a so-called Fermi-Einstein condensate.
This mechanism is further illustrated in the Schwinger model where it
solves the Silver-Blaze problem.}

\FullConference{International Workshop on QCD Green's Functions, Confinement and Phenomenology,\\
		September 05-09, 2011\\
		Trento Italy}

\begin{document}

\section{Introduction:}

Under normal conditions, eventually the most prominent feature of
hadronic matter is colour confinement: quarks merely act as
constituents of hadrons and qualify as auxiliary fields in the sense that
the QCD partition function is solely given in terms of states with
$N$-ality zero. Supposedly, the situation changes under extreme conditions,
temperature and/or density, where quarks and gluons are liberated.
This picture is corroborated at zero density and high temperatures by means
of lattice Monte-Carlo simulations, which offer a first principle
approach with controllable error margins. The picture is clear-cut in the
heavy quark limit: the hadronic phase is characterised by a Wigner-Weyl
realisation of the centre symmetry while the so-called gluon plasma phase
at high temperature relates to a spontaneously breaking of this
symmetry~\cite{Svetitsky:1982gs,Svetitsky:1985ye}. Quite
recently~\cite{arXiv:1109.0502,arXiv:1111.4315}, it was argued that this
picture extends to the theory with
dynamical quarks as well: although centre symmetry is explicitly broken
by the quark matter, centre sector transitions do still take place in the
hadronic phase~\cite{arXiv:1109.0502} (see below for further details).
The role of the quarks here is to merely induce a
bias  towards the trivial centre sector. Only at high temperature, the
centre symmetry also breaks spontaneously giving rise to the
quark gluon plasma phase.

\vskip 0.3cm
The properties of cold, but dense QCD matter is far less understood due
to the infamous sign-problem which prevents Monte-Carlo simulations
based upon Importance Sampling. A promising attempt to reach beyond the
scope of  Importance Sampling and to simulate QCD in this region of the phase
diagram resorts to Complex Langevin dynamics~\cite{200293} though further
studies are needed to understand their convergence
properties~\cite{arXiv:0912.3360,arXiv:0912.0617}. The intuition on
dense fermionic matter arises from the free Fermi gas model:
Antiperiodic boundary conditions in (Euclidean) time direction imply
the formation of a Fermi sphere for finite chemical potential.
The spin-statistics theorem~\cite{45434,weinberg1} ties particles with
half-integer spin to fermion statistics. Note, however, that the theorem
requires  propagating (finite mass) particles with a positive definite norm.
The statistics of {\it confined} quarks is therefore not necessarily
restricted by the spin-statistics theorem. In fact, it was shown
in~\cite{Langfeld:2009cd} that in the confinement regime
a certain ``large'' change of
the gluon background  field can be traded in for a change of the quark
boundary conditions~\cite{Langfeld:2009cd}. Another well known example
originates from  QCD perturbation theory: fermionic ghost fields inherit
periodic boundary conditions from the gluon sector and evade the
spin-statistics theorem  since they involve negative norm states.
The observation that quarks do not necessarily comply with anti-periodic
boundary conditions, but, at least for an even number of colours,
might acquire Bose statistics has a tremendous phenomenological
impact: if, by virtue of the gluonic background, quarks develop a
Matsubara zero mode, they might underdo condensation if the
chemical potential equals the quark mass gap. This phenomenon has been
called {\it Fermi-Einstein
  condensation}~\cite{arXiv:0911.0319,arXiv:1109.0502,arXiv:1111.4315}.

\vskip 0.3cm
Considering confined quarks as auxiliary fields, a constraint to
thermodynamical observables emerge form the sheer fact that
the sole excitations are provided by hadrons: at zero temperature,
observables are independent of the chemical potential as long as
it is smaller than the hadronic mass gap. A widespread problem is that
approximate methods and QCD models fail to comply with this physical fact.
Accordingly, this problem has been called the {\it Silver Blaze} problem
by Cohen in~\cite{Cohen:2003kd} reminiscent of the corresponding short
story by Sir Arthur Conan Doyle, in which a dog doing nothing provides the
essential clue to solve the crime mystery.


\section{Centre sector transitions }

\begin{figure}
  \includegraphics[width=10cm]{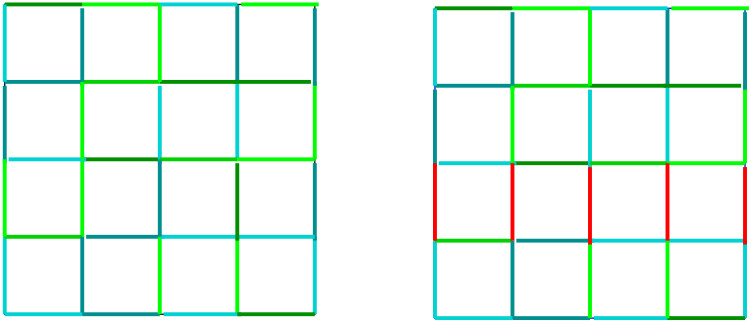}
\caption{\label{fig:1} Sketch of a centre sector transition from
a generic gluonic lattice configuration (left) to its centre
transform (right).
}
\end{figure}
Let us adopt the lattice notation to illustrate the impact of centre
sector transitions. A generic gluon configuration is given in terms
of the links $U_\mu (x) \in SU(N_c)$ and is illustrated in figure~\ref{fig:1},
left panel. A particular Polyakov line is defined by
\be
P(\vec{x}) \; = \; \frac{1}{N_c} \, \tr \, \prod _{x_0=1}^{N_c} U_0 (x_0,
\vec{x}) \; .
\label{eq:1}
\en
A centre transformed configuration is generated by multiplying
the time-like links of a particular time slice $t$ by a centre element
$Z_n = z_n \, 1 \in SU(N_c)$:
\be
U^{(n)}_0(t,\vec{x}) \; = \; Z_n \, U_0(t,\vec{x}) \; , \; \; \; \;  \forall \;
\vec{x} , \hbo U^{(n)}_\mu(x) \; = \; U_\mu(x)  \; \; \; \; \hbox{else}.
\label{eq:2}
\en
It is easy to see that any lattice action $A$ of pure Yang-Mills theory, which
is built upon closed contractible loops, is invariant under the
transformation (\ref{eq:2}) while the Polyakov loop is not:
$$
A[ U_\mu ] = A\left[U_\mu ^{(n)} \right] , \hbo
P \left[ U^{(n)} _\mu \right](\vec{x}) \; = \; z_n \; P[ U_\mu ](\vec{x})
\hbo  z_n = \exp \left\{ i \, \frac{2\pi}{N_c} n \right\} , \; \; \;
n=1 \ldots N_c \; .
$$
Using the spatial Polyakov loop average $p$, it is straightforward to
associate a so-called centre sector to each lattice configuration:
\be
p = \frac{1}{V} \sum _{\vec{x}} P (\vec{x}), \hbo
C(p) = n , \hbo n: \; \Big\vert \; \mathrm{arg} (p)
- \frac{2 \pi n }{N_c}  \, \Big\vert \to \mathrm{min} \; ,
\label{eq:3}
\en
where
$$
\mathrm{arg} (p) = \varphi \in \,]0, 2\pi] \; , \hbo
p \; = \; \vert p \vert \, \exp \{ \mathrm{i} \varphi \} \; .
$$
It is straightforward to show that the centre transformation (\ref{eq:2})
shifts the centre sector by $n$:
$$
C\left( p\left[U_\mu \right] \right) \; = \; m \hbo \rightarrow \hbo
C\left( p\left[U^{(n)} _\mu \right] \right) \; = \; m \; + \; n \; .
$$
It is the symmetry which is spanned by the transformation (\ref{eq:2})
which is {\it spontaneously broken} in high temperature pure Yang-Mills
theory~\cite{Svetitsky:1982gs,Svetitsky:1985ye}. The phenomenon
related to this breakdown is {\it deconfinement}: the
static quark antiquark potential is no longer linearly rising at
large distances, and the gluons contribute black body radiation to the
thermal energy density suggesting gluon liberation.

\vskip 0.3cm
\begin{figure}
  \includegraphics[width=6cm]{tunn.eps}
  \includegraphics[width=9cm]{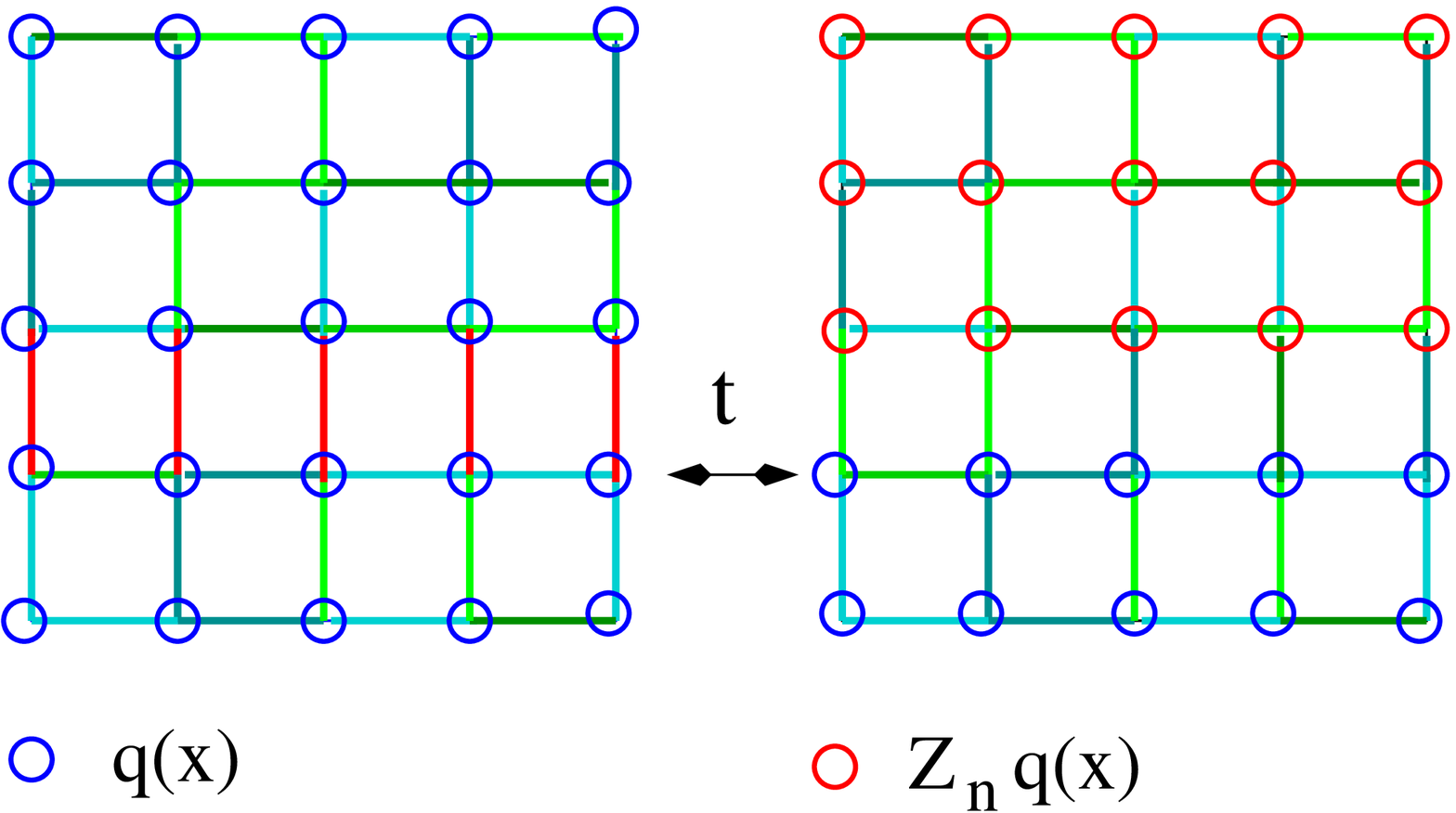}
\caption{\label{fig:2} Right: Sector transition probability $\tau $
for the pure SU(2) YM-theory and the SU(2)-Higgs theory. Left:
Roberge-Weisz transformation undoing a gluonic centre sector transition.
}
\end{figure}
How does this picture change for QCD-like theories, i.e., for gauge theories
with dynamical matter which transforms under the fundamental representation
of the gauge group? Let us consider for the moment $SU(N_c)$ Yang-Mills
equipped with a quark determinant arising from integrating over the
dynamical quark fields. The transformation (\ref{eq:2}) is no longer a
symmetry of the action since the quark determinant provides a bias
towards the trivial centre sector $n=N_c$. The question here is whether
the lattice configurations are confined to the trivial centre sector
because of this {\it explicit} centre symmetry breaking:
$$
C\left( p\left[U_\mu \right] \right) \; \stackrel{?}{=} \; N_C \; , \hbo
U_\mu \; \stackrel{?}{\longrightarrow\kern-1.2em\slash\kern+1em} \;
U^{(n)}_\mu \; , \; \; \; \; \; n \; =\kern-.9em\slash \; \; N_c \hbo
\hbox{(YM-theory with matter).}
$$
This question was thoroughly studied in~\cite{arXiv:1109.0502} for the
SU(2)-Higgs gauge theory for which the Higgs field plays the role of the
dynamical matter. To detect whether centre sector transitions do
occur, we divide the spatial lattice universe into two parts and calculate
the spatially averaged Polyakov line over each part. Let us call the
result $p_<$ and $p_>$. Equally well, we can associate a centre sector
to each half of the lattice universe by the mapping
$C(p_<)$  and $C(p_>)$. We then consider the probability $\tau $ that
both centre sectors are different, i.e., $C(p_<) =\kern-.9em\slash \;
C(p_>)$. If centre sector transitions do occur, $C(p_{>})$ and $C(p_{<})$
roughly sample all sectors ranging from $1 \ldots N_c$ leaving us with
$\tau \approx 1- 1/N_c$. If, on the other hand, centre sector transitions
are prohibited, we would have $\tau =0$. The so-called ``tunnelling
coefficient'' $\tau $ is shown in figure~\ref{fig:2}, right panel, for the
pure SU(2) Yang-Mills theory as function of the Wilson-$\beta $ parameter.
While $\tau $ is close to $1/2$ at low temperatures, it rapidly drops
for large $\beta $ values, which correspond to the gluon plasma phase.
Also shown are the findings for the SU(2) Higgs theory: note that
$\tau $ is still close to $1/2$ at small $\beta $ indicating that
centre sector transitions do frequently occur. It is only above a
certain critical value when these transitions cease to exist due to
a spontaneous breaking of the centre symmetry besides of its explicit breaking.
Also note that the critical $\beta $ value for deconfinement is smaller
for the Higgs theory than for the pure YM-theory. This shows that
the deconfinement critical temperature is smaller for the Highs theory,
which is expected due to the matter effects.

\vskip 0.3cm
Let us now include dynamical quark matter $q(x) $ which is subjected to a
centre transformed gluon background field $U_\mu ^{(n)}$ (\ref{eq:2})
(see figure~\ref{fig:2}, right panel). The matter gluon interaction
is assumed to be of next-to-nearest-neighbour type:
$\bar{q}(x) \, U_\mu (x) \, q(x+\mu )$. It turns out~\cite{Langfeld:2009cd}
that the centre transformation of the gluon fields can be reversed by
transforming the quarks fields via
\be
q^{(n)}(x) \; = \; Z_n \; q(x) \; \; \; \hbox{for} \; \; x_0>t, \hbo
q^{(n)}(x) \; = \; q(x) \; \; \; \hbox{else.}
\label{eq:4}
\en
This {\it almost } appears to be just a change of variables and therefore
an invariance of the partition function. Note, however,
that with the transformation (\ref{eq:4}) the boundary conditions
of the quarks change from antiperiodic to $Z_n$-periodic:
$$
q(x_0 1/T,\vec{x}) \; = \; (-1) \; Z_n \; q(x_0,\vec{x})
$$
Let us $ \Det _{\mathrm{AP} } $ denote the quark determinant with antiperiodic
boundary conditions and let $\Det _{(n) } $ represent the determinant
with $Z_n$-periodic boundary conditions. Since the  partition function of QCD
(or the QCD-like theory) is a gluonic ensemble average over all centre
sectors, the partition function can be written as
\be
\int {\cal D} U_\mu \; \Det _{\mathrm{AP} } [ U_\mu ] \;
\exp \{ -S_\mathrm{YM}[U_\mu] \}  \; = \;
\int {\cal D} U^{(N_c)}_\mu  \; \Biggl( \sum _n \Det _{(n) } \Bigl[ U^{(N_c)}_\mu]
\, \Biggr) \; \exp \left\{ - S_\mathrm{YM}[U^{(N_c)}_\mu]
\right\}  \; .
\label{eq:5}
\en
Note that for an even number of colours, there is the centre element
$Z_{N_c/2} = -1$. Hence, there is a particular centre sector, which
gives rise to {\it periodic} boundary conditions for the quark
determinant when the gluon fields are mapped to the trivial sector
by means of the transformation (\ref{eq:4}). In zeroth order perturbation
theory, i.e., for $ U^{(N_c)}_\mu=1$, the right hand side of (\ref{eq:5})
given rise to a centre-symmetric Fermi gas model. For $N_c$ even, it was
shown in~\cite{Langfeld:2009cd} that when the chemical potential $\mu $
approaches the mass gap $m$, the baryon density $\rho $ is largely
determined by the centre sector $n=N_c/2$ alone:
\be
\rho \; \approx \; \int_m dE \; \frac{ -1 }{ \mathrm{e}^{[E-\mu]/T } -1 } \; ,
\hbo \hbox{ ($N_c$ even, confinement phase). }
\label{eq:6}
\en
where $E$ can be interpreted as the one-particle energy of the
(modified) Fermi gas model. Notably, this contribution is singular
for $\mu \to m$, and suggests an instability due to condensation
quite analogously to Bose-Einstein condensation. Since the degrees
of freedom which condense are fermions which are exposed to a non-trivial
centre background field, this has been called {\it Fermi-Einstein
  condensation} (FEC). It is important to note that FEC can only occur
in the confined phase: in this phase, transitions from the trivial
centre sector $n=N_c$ to the sector with $n=N_c/2$ do occur. Under extreme
conditions, this is no longer the case: the sector transitions cease to exist,
and only the trivial centre sector background is attained. In this case,
the above model coincides with the standard Fermi gas model
displaying Fermi statistics:
\be
\rho \; \approx \; \int_m dE \; \frac{ 1 }{ \mathrm{e}^{[E-\mu]/T } +1 } \; ,
\hbo \hbox{ (high temperature deconfinement phase). }
\label{eq:7}
\en
The question arises whether FEC also takes place in QCD-like theories
with an {\it odd} numbers of colours and, most importantly, in QCD.
To gain first insights, the centre symmetric quark model, i.e.,
zeroth order perturbation theory, has been generalised to the SU(3)
gauge group~\cite{arXiv:1109.0502}. The phase diagram can be analytically
calculated in this model. It was found that, for low temperature and
intermediate values of the chemical potential, a phase with an excess
of baryon density does occur if the spatial volume is small enough, i.e.,
if the system is under pressure~\cite{arXiv:1109.0502}.

\section{Lessons from the Schwinger model }

The Schwinger model~\cite{Schwinger:1962tp}, i.e., QED in two dimensions on a
space-time torus, is an ideal testbed for tracing out new ideas since many
computations can be done analytically. Here, we will study the phenomenology
of the centre-sector transitions in this model for a non-vanishing chemical
potential.

\vskip 0.3cm
The model with massless fermions was exactly solved in Hamiltonian formalism
on the line in~\cite{Brown:1963,Lowenstein:1971fc,Casher:1974vf}
and on $S^1$ in \cite{Manton:1985jm,Iso:1988zi}. The model on the torus has
been studied in \cite{Joos:1990km} and in particular the temperature
dependence  of the chiral condensate, Wilson loop correlators and Polyakov
line correlators have been
determined~\cite{Sachs:1992pa,Smilga:1992hx,Azakov:1996xk}.
In turns out that chiral symmetry is anomalously broken and that only states
with a vanishing net baryon number appear in the spectrum.
The latter observation is key and implies that the baryon density ought to
vanish even in the case of non-vanishing values of the chemical potential.
Non-vanishing values of the fermion chemical potential have been
firstly considered in~\cite{Sachs:1995dm,AlvarezEstrada:1997ja}, and
it was found, indeed, that the full non-perturbative partition function is
independent of the chemical potential.

\vskip 0.3cm
The gauge potential can be written as~\cite{Sachs:1992pa}
\be
A_0=\frac{2\pi}{\beta}h_0+\partial_0\lambda-\partial_1\phi,\quad
A_1=\frac{2\pi}{L}h_1+\partial_1\lambda+\partial_0\phi
\label{eq:10}
\en
where the periodic functions $\lambda$ and $\phi$ integrate to zero.
The constant toron fields $h_{0,1} \in [0,1[$ label the U(1) centre sectors
of the model. This can be easily seen from the fact that the shift
$h_0 \to h_0 + \alpha $ transforms the Polyakov line by a U(1) centre
element $ P(x) \; \to \; \exp \{ \mathrm{i} \, 2\pi \, \alpha \} \; P(x)$.
The partition function factorise into a photonic part
and into the centre sector average of the fermion determinant:
\be
Z(T,L,\mu) =
(2\pi)^2 \sqrt{\frac{{\det}'(-\triangle)}{{\det}'(-\triangle+m_\gamma^2)}}
\; \int_0^{1} dh_0 \; dh_1 \, \det(\mathrm{i}\slashed{\partial}_{h,\mu})\,,
\label{eq:11}
\en
where $m_\gamma $ is a dynamically generated photon mass, and $L$ is the
spatial extent of the torus. {\it Assuming} a frozen centre sector, i.e.,
a fixed value for $h_0$, would lead to a {\it Silver Blaze} problem since
the baryon density $\rho_B$ is non-vanishing in this case:
$$
\rho_B \stackrel{L\to\infty}{\longrightarrow}
\frac{1}{\pi}\int_0^\infty dp\left\{
\frac{z}{ \mathrm{e}^{\beta (p-\mu)}+z}-
\frac{z^\ast }{ \mathrm{e}^{\beta (p+\mu)}+ z^\ast}\right\} \; , \hbo
z=\exp \{ - 2 \pi \mathrm{i} \, h_0 \}.
$$
Note also that for the particular choice $h_0=1/2$, we find $z=-1$ implying
that the one-particle distribution functions are of Bose type.
Thus, freezing the centre sector yields the wrong physics.
On the other hand, the centre sector average, i.e., the integration
over the toron field $h_0$, yields a result which is independent of
$\mu $~\cite{arXiv:1109.0502}, i.e.,
$$
Z(T,L,\mu) \; = \; \sqrt{\frac{V}{2}}\frac{1}{
\sqrt{{\det}'(-\triangle+m_\gamma^2)}} \; ,
$$
and, thus, solves the Schwinger-Blaze problem in the Schwinger model.


\begin{thebibliography}{99}
\bibitem{Svetitsky:1982gs}
  B.~Svetitsky, L.~G.~Yaffe,
  Nucl.\ Phys.\  {\bf B210 } (1982)  423.

\bibitem{Svetitsky:1985ye}
  B.~Svetitsky,
  Phys.\ Rept.\  {\bf 132 } (1986)  1-53.

\bibitem{arXiv:1109.0502}
  K.~Langfeld and A.~Wipf,
  {\it Fermi-Einstein condensation in dense QCD-like theories},
  arXiv:1109.0502 [hep-lat], accepted by Annals of Physics.

\bibitem{arXiv:1111.4315}
  K.~Langfeld and A.~Wipf,
  {\it From confinement to new states of dense QCD matter},
  arXiv:1111.4315 [hep-lat], in press by Prog.~Part.~Nucl.~Phys.

\bibitem{200293}
  G.~Parisi,
  Phys.\ Lett.\ B\ {\bf 131} (1983) 393.

\bibitem{arXiv:0912.3360}
  G.~Aarts, E.~Seiler and I.~-O.~Stamatescu,
  Phys.\ Rev.\ D\ {\bf 81} (2010) 054508
  [arXiv:0912.3360 [hep-lat]].

\bibitem{arXiv:0912.0617}
  G.~Aarts, F.~A.~James, E.~Seiler and I.~-O.~Stamatescu,
  Phys.\ Lett.\ B\ {\bf 687} (2010) 154
  [arXiv:0912.0617 [hep-lat]].

\bibitem{45434}
  W.~Pauli,
  Phys.\ Rev.\ \ {\bf 58} (1940) 716.

\bibitem{weinberg1}
Steven Weinberg,
{\it The Quantum Theory of Fields: Volume 1, Foundations},
Cambridge University Press 1995.

\bibitem{Langfeld:2009cd}
  K.~Langfeld, B.~H.~Wellegehausen and A.~Wipf,
  Phys.\ Rev.\  D {\bf 81} (2010) 114502
  [arXiv:0906.5554 [hep-lat]].

\bibitem{arXiv:0911.0319}
  K.~Langfeld,
  PoSQCD\ {\bf -TNT09} (2009) 022
  [arXiv:0911.0319 [hep-lat]].

\bibitem{Cohen:2003kd}
  T.~D.~Cohen,
  Phys.\ Rev.\ Lett.\  {\bf 91} (2003) 222001
  [arXiv:hep-ph/0307089].

\bibitem{Schwinger:1962tp}
  J.~S.~Schwinger,
  Phys.\ Rev.\  {\bf 128 } (1962)  2425-2429.

\bibitem{Brown:1963}
L.S.~Brown,
Nuovo Cimento {\bf 29} (1963) 617-643

\bibitem{Lowenstein:1971fc}
J.~Lowenstein, A.~Swieca. Ann.\ Phys. {\bf 68} (1971) 172-195;

\bibitem{Casher:1974vf}
A.~Casher, J.~B.~Kogut, L.~Susskind,
Phys.\ Rev.\  {\bf D10 } (1974)  732-745.

\bibitem{Manton:1985jm}
N.~Manton, Ann.\ Phys. {\bf 159} (1985) 220-251


\bibitem{Iso:1988zi}
S.~Iso, H.~Murayama,
Prog.\ Theor.\ Phys.\  {\bf 84 } (1990)  142-163.

\bibitem{Joos:1990km}
H. Joos,
Helv.~Phys.~Acta {\bf 63} (1990) 670-682.

\bibitem{Sachs:1992pa}
I.~Sachs and A.~Wipf,
Helv.\ Phys.\ Acta\ {\bf 65} (1992) 652-678.
[arXiv:hep-th/1005.1822]

\bibitem{Smilga:1992hx}
  A.~V.~Smilga,
  Phys.\ Lett.\  {\bf B278 } (1992)  371-376.

\bibitem{Azakov:1996xk}
  S.~Azakov,
  Fortsch.\ Phys.\  {\bf 45 } (1997)  589-626.
  [hep-th/9608103].

\bibitem{Sachs:1995dm}
I.~Sachs and A.~Wipf,
Annals Phys. {\bf 249} (1996) 380-429.
[arXiv:hep-th/9508142]


\bibitem{AlvarezEstrada:1997ja}
  R.~F.~Alvarez-Estrada, A.~Gomez Nicola,
  Phys.\ Rev.\  {\bf D57} (1998) 3618-3633.
  [hep-th/9710227].


\end{thebibliography}
\end{document}